# An omnidirectional shear horizontal wave transducer based on ring array of face-shear ($d_{24}$) piezoelectric ceramics

Hongchen Miao[1,2], Qiang Huan[1], Qiangzhong Wang[1], Faxin Li[1,2, a]

[1]LTCS and Department of Mechanics and Engineering Science, College of Engineering, Peking University, Beijing, 100871, China

[2]Center for Applied Physics and Technology, Peking University, Beijing, 100871, China

**Abstract**

The non-dispersive fundamental shear horizontal ($SH_0$) wave in plate-like structures is of practical importance in non-destructive testing (NDT) and structural health monitoring (SHM). Theoretically, an omnidirectional $SH_0$ transducer phased array system can be used to inspect defects in a large plate in the similar manner to the phased array transducers used in medical B-scan ultrasonics. However, very few omnidirectional SH transducers have been proposed so far. In this work, an omnidirectional SH wave piezoelectric transducer (OSH-PT) was proposed which consists of a ring array of twelve face-shear ($d_{24}$) trapezoidal PZT elements. Each PZT element can produce face-shear deformation under applied voltage, resulting in circumferential shear deformation in the OSH-PT and omnidirectional SH waves in the hosting plate. Both finite element simulations and experiments were conducted to examine the performance of the proposed OSH-PT. Experimental testing shows that the OSH-PT exhibits good omnidirectional properties, on matter it is used as a SH wave transmitter or a SH wave receiver. This work may greatly promote the applications of SH waves in NDT and SHM.

**Keywords**: shear horizontal waves; guided waves; face-shear piezoelectrics; non-destructive testing (NDT), structural health monitoring (SHM).

---

a) Author to whom all correspondence should be addressed, Email: lifaxin@pku.edu.cn.

## 1. Introduction

Ultrasonic guided-wave-based inspection technique has been proven to be a very effective method for defects inspecting in large waveguide structures [1]. For example, an omnidirectional guided wave transducer phased array system can be used to inspect all the surrounding area of the plate up to a distance of several meters[2], since wave energy can be focused at any target direction and the omnidirectional B-scan can be ultimately performed[3]. Such phased array system requires that each individual array element should be also an omnidirectional transducer, with equal transmission and reception sensitivity of the chosen wave mode in all directions[3]. In the past decades, various omnidirectional Lamb waves transducers have been proposed, based on electromagnetic acoustic principle[4-6], magnetostrictive effect[7] or piezoelectric effect[8-10].

Compared with the dispersive lamb waves, the non-dispersive fundamental shear horizontal ($SH_0$) wave in plate-like structures or the fundamental torsional (T(0,1)) wave in pipe-like structures is more promising in practical applications[11]. Besides non-dispersion, $SH_0$ wave will not convert to Lamb waves when encountered with defects or boundaries, thus reducing the complexity of the received wave signals[12]. Furthermore, the attenuation of the $SH_0$ (or T(0,1)) wave is theoretically zero when the waveguide is surrounded by non-viscous liquid such as water [2]. In spite of these attractive features of the $SH_0$ wave, so far very limited omnidirectional $SH_0$ wave transducers have been proposed. Seung et al developed a magnetostrictive electromagnetic acoustic transducer(EMAT) for exciting and measuring omnidirectional $SH_0$ wave[13]. Wei et al proposed an omnidirectional $SH_1$ wave EMAT which is also based on the magnetostrictive effect[14]. Very recently, Seung et al proposed a new omnidirectional $SH_0$ wave EMAT based on the Lorentz force [15]. Obviously, these omnidirectional $SH_0$ wave EMATs can only be used for NDT in metallic structures and are not suitable for structural health monitoring (SHM) due to their relative large size and low energy conversion efficiency.

Piezoelectric transducers are more promising in the fields of both NDT and SHM, due to their compact size and peculiar electromechanical coupling properties. In most cases, piezoelectric acoustic transducers are used in the $d_{33}$ mode (normal probe for scanning) or the $d_{31}$ mode (such as in exciting/receiving Lamb waves in plates). In recent years, there are some works on excitation/reception of SH waves (or torsional waves) based on piezoelectric shear mode transducers, most of which employed the thickness-shear ($d_{15}$) mode [16, 17] and a few adopted the face-shear ($d_{36}$ or $d_{24}$) mode[12, 18, 19]. All these piezoelectric SH transducers have strong directivity with one or two major directions, which is not suitable for sensor applications in SHM. So far, only one omnidirectional SH wave piezoelectric transducer has been proposed, by Belanger and Boivin [20] based on a circular array of six $d_{15}$ mode PZT patches. Their SH wave transducer can generate omnidirectional mixed-mode SH wave and Lamb wave, while the reception capacity was not demonstrated[20]. Actually, the omnidirectional properties are more important for wave receivers than for wave transmitters. Thus, it is necessary to develop an omnidirectional piezoelectric SH wave transducer which can behave as both transmitter and receiver. On the other hand, it has been shown in our recent work[19] that the face-shear piezoelectrics is superior to thickness shear piezoelectrics in driving SH waves because of its higher efficiency of energy transfer. Meanwhile, recently we have realized the face-shear $d_{36}$ mode in PbZr$_{1-x}$Ti$_x$O$_3$ (PZT) ceramics [21, 22]and successfully excited SH$_0$ wave in an aluminum plate using $d_{36}$ mode PZT wafers[12]. Very recently, we proposed a face-shear ($d_{24}$) PZT transducer and successfully excited/received pure SH$_0$ wave over a wide frequency range[19]. Therefore, it is promising to develop an omnidirectional SH$_0$ wave transducer with higher energy conversion efficiency by using the face-shear piezoelectrics.

In this work, we proposed an omnidirectional SH wave piezoelectric transducer, which can excite and receive pure SH wave over full 360° in plate-like structures. The proposed omnidirectional transducer is made up of a ring array of twelve face-shear ($d_{24}$) mode PZT elements. Firstly, the configuration and working principle of the proposed omnidirectional transducer is presented. Then finite element simulations were performed to investigate the omni-directivity of the SH$_0$ wave excited by the proposed transducer. Finally, experiments were conducted to examine the performance of exciting/receiving the desired SH$_0$ wave and the omni-directivity of the proposed transducer. For convenience, the proposed piezoelectric transducer is referred to as OSH-PT (omnidirectional SH wave piezoelectric transducer) in the subsequent sections.

## 2. Configuration and working principle of the OSH-PT

As we know, a point transducer that can cause and measure in-plane particle vibrations could be used to excite and detect SH wave propagating perpendicular to the direction of particle's vibration (not in an omnidirectional manner)[23]. In order to excite the omnidirectional SH wave, a particular excitation force with axisymmetric distribution and polarized in the tangential direction is required as shown in Fig. 1(a)[2]. To realize such a particular excitation force, the proposed OSH-PT is designed consisting of several identical face-shear $d_{24}$ trapezoidal PZT elements with the thickness of 1.5 mm, as shown in the Fig. 1(b). The height "ae" of the isosceles trapezoid was set to be half wavelength of the target $SH_0$ wave at a given excitation frequency, ensuring the largest shear stresses appear on the upper line and baseline of the isosceles trapezoid, respectively. In order to optimize the face-shear performance of the trapezoidal PZT elements, the baseline "dc" of the isosceles trapezoid was set to be equal to its height "ae". As shown in Fig. 1(b) and (c), each $d_{24}$ PZT element is in-plane poled along the "3" direction, thus the effective polarization of the OSH-PT tends to a circumferential polarization. Obviously, if the angle θ between the two waists of the trapezoidal PZT elements becomes smaller, the effective circumferential polarization will be more uniform, but the size of the OSH-PT will become larger. Our analysis shows that $\theta \approx 30°$ is an appropriate value, rendering the number of the PZT elements to be twelve. The photo of actually-fabricated ring-shaped OSH-PT is shown in Fig. 1(d). Now Let us explain how the circumferential shear stress could be produced. As shown in Fig. 1(c), the driving field of the trapezoidal PZT elements is along the "2" direction and the thickness is along the "1" direction. When voltage is applied to the trapezoidal PZT elements, pure face-shear deformation is expected to be obtained due to the face-shear piezoelectric $d_{24}$ mode. Thus each trapezoidal PZT element can be regarded as shear point source in the OSH-PT and an effective circumferential shear stress is expected to be generated.

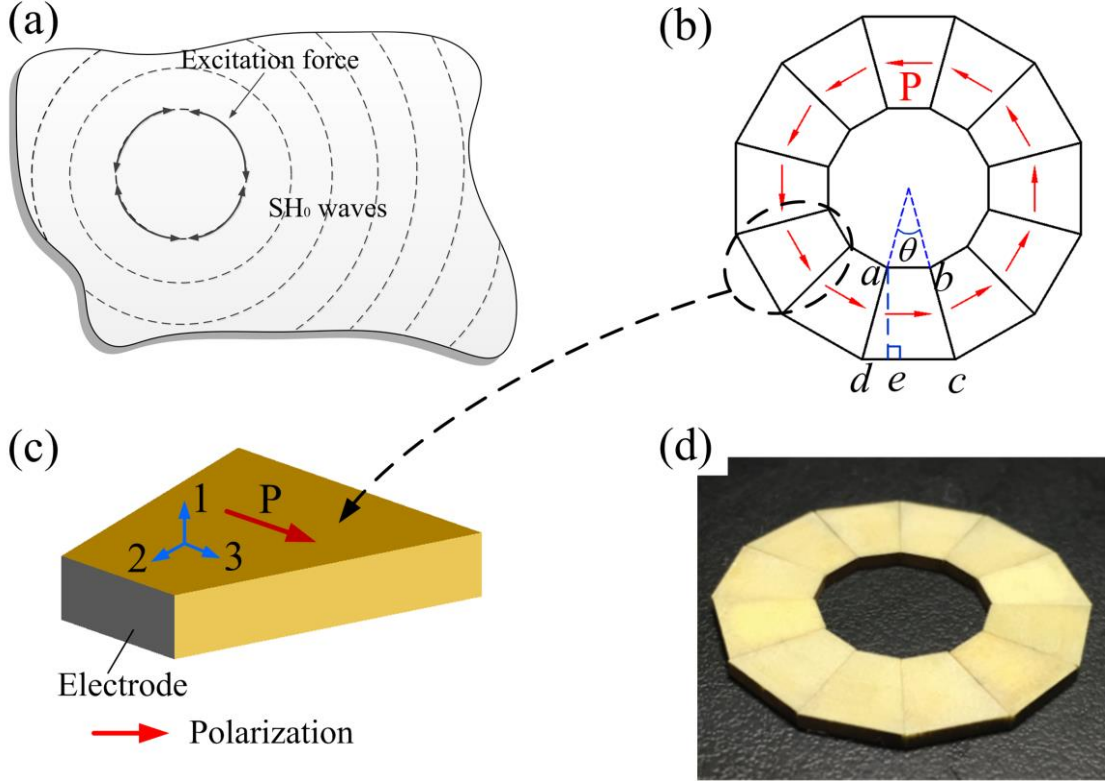

**Figure 1** (a) Schematic of the excited $SH_0$ wave field by distributed excitation force polarized in the circumferential direction. (b) Schematic top view of the proposed OSH-PT. (c) Schematic of the face-shear ($d_{24}$) mode trapezoidal PZT element for fabricating OSH-PT. (d) Photo of the fabricated OSH-PT.

**3. Finite element simulations**

Firstly, finite element (FEM) simulations based on ANSYS were performed to predict the generation of the desired $SH_0$ wave mode and the omni-directivity of the proposed OSH-PT. An aluminum plate with the dimensions of 400 mm×400 mm×1 mm was used in the simulation. Its Young's modulus, Poisson ratio and density were 70 GPa, 0.33 and 2700 kg/m$^3$, respectively. An OSH-PT with the desired working frequency of 190 kHz was used in the simulation. Therefore, the height "ae" and baseline "dc" of the trapezoidal PZT elements were set to be 8 mm (approaching the half wavelength of the desired $SH_0$ wave). The θ between the two waists of trapezoidal PZT elements was 30° and twelve elements were used. The material parameters of the PZT elements can be found in [24] and will not be listed here. The proposed OSH-PT was boned on the aluminum plate using a thin bond layer with the elastic modulus of 500 MPa and thickness of 50 μm. In the simulation, the aluminum plate was modeled by SOLID185 elements and the OSH-PT was modeled by SOLID5 elements in the ANSYS software. A five-cycle Hanning window-modulated sinusoid toneburst was used to excite the proposed OSH-PT. The amplitude of the driving signal was set to be 20 V and its central frequency varied from 160 kHz to 220 kHz to investigate the frequency characteristics of the proposed OSH-PT.

Fig. 2 shows the FEM simulated displacement wavefields excited by the proposed OSH-PT at 190 kHz. As the particle vibration caused by the $SH_0$ wave is in-plane and perpendicular to the wave propagating direction, the tangential displacement component in the cylindrical coordinates can represent the $SH_0$ wave. Similarly, the radial displacement component and out-of-plane displacement component in the cylindrical coordinates can be used to roughly represent the $S_0$ wave mode and $A_0$ wave mode respectively. As expected, Fig. 2(a) shows that the generated tangential displacement component is axisymmetric around the transducer, implying the good omnidirectional property of the $SH_0$ waves excited by the proposed OSH-PT. Meanwhile, it was found that the amplitude of the tangential displacement component is about one order higher than that of the radial and out-of-plane displacement components, as shown in Fig. 2(a)-(c). From Fig. 2(a) and (d), it can be seen that the tangential displacement component occupies about 99.5% of the total displacement. These phenomena indicate that pure $SH_0$ wave mode is expected to be excited by the proposed OSH-PT, since the amplitudes of the $S_0$ and $A_0$ wave modes are negligible.

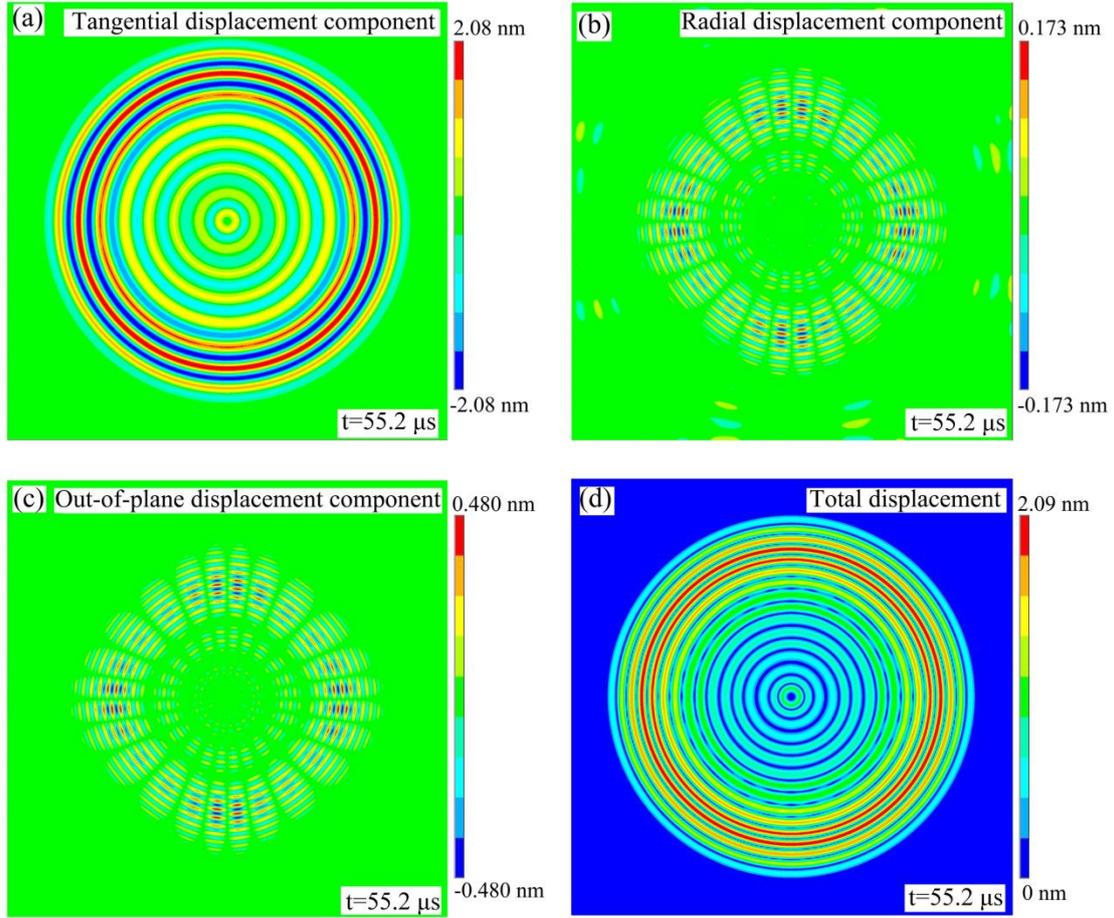

**Figure 2** Finite element method (FEM) simulated displacement wavefields excited by the proposed OSH-PT at 190 kHz: (a) tangential displacement component ($SH_0$ wave), (b) radial displacement component ($S_0$ wave) and (c) out-of-plane displacement component ($A_0$ wave) in the cylindrical coordinates, (d) total displacement.

To further explore the omni-directivity of the proposed OSH-PT, the simulated tangential displacement components caused by the excited $SH_0$ waves at 190 kHz were extracted at the interval of 15° from 0° to 180° and the results were plotted in Fig. 3(a). As expected, the obtained amplitude spectrums of the excited $SH_0$ waves with different propagating directions are identical, indicating the proposed OSH-PT exhibits perfect omni-directivity properties. Fig. 3(b) shows the amplitude directivity of the $SH_0$ waves excited by the proposed OSH-PT at 160 kHz, 190 kHz and 220 kHz, respectively. The plots are normalized against the excited $SH_0$ wave amplitude at 190 kHz. It can be seen that the OSH-PT exhibits perfect omni-directionality in this frequency range (160 kHz to 220 kHz). Also as expected, the wave amplitude reaches its maxima at 190 kHz. The amplitude can keep about 80% of the maxima at 220 kHz, while it quickly decreases to about 64% of the maxima at 160 kHz. These results confirmed the necessity of the size design for the OSH-PT.

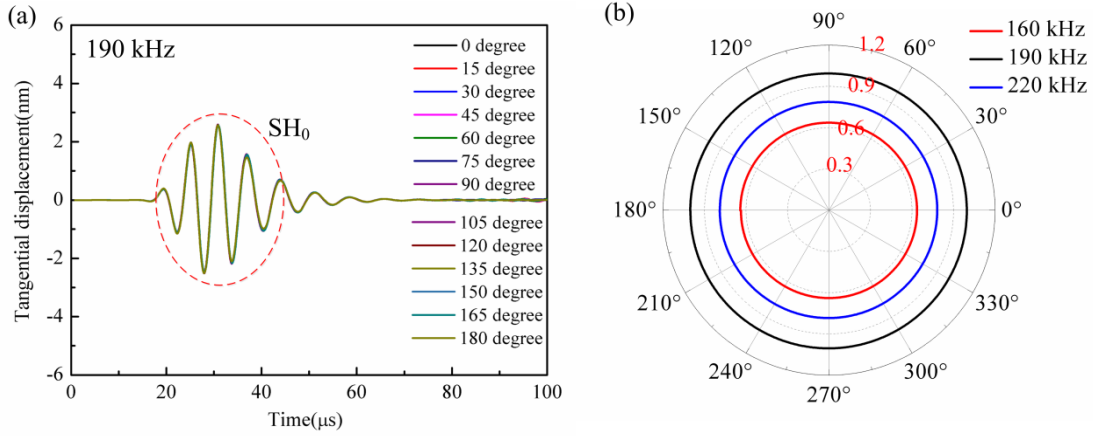

**Figure 3** (a) The FEM simulated amplitude spectrums of the excited $SH_0$ waves with different propagating directions at 190 kHz; (b) The simulated amplitude directivity of the $SH_0$ waves excited by the proposed OSH-PT at 160 kHz, 190 kHz and 220 kHz. (The plots are normalized against the excited $SH_0$ wave at 190 kHz.)

## 4. Experimental validation

Experiments were then performed to examine the proposed OSH-PT's performance on excitation and reception of omnidirectional $SH_0$ waves. The OSH-PT used in the experiments has the same size and material parameters used in the FEM simulations in Section 3. An aluminum plate with the dimensions of $2000 \text{ mm} \times 1000 \text{ mm} \times 1 \text{ mm}$ was used in the testing. Firstly, the OSH-PT served as the transmitter to excite $SH_0$ wave. A $d_{36}$ face-shear PMN-PT crystal patch($d_{36}$=1600 pC/N and $d_{31}$=-360 pC/N) with dimensions of $5 \text{ mm} \times 5 \text{ mm} \times 1 \text{ mm}$ was used as the receiver to check the purity of the excited $SH_0$ wave, since it can receive both $SH_0$ wave and Lamb waves[12, 18]. Then, to check whether the OSH-PT could receive $SH_0$ wave, a face-shear $d_{24}$ PZT-5H wafer ($6 \text{ mm} \times 6 \text{ mm} \times 1.5 \text{ mm}$) which can generate pure $SH_0$ wave, was used as the transmitter. Detailed properties of the $d_{24}$ type PZT-5H wafer can be found in our recent work[19]. Besides the OSH-PT, a $d_{36}$ type PMN-PT crystal wafer ($5 \text{ mm} \times 5 \text{ mm} \times 1 \text{ mm}$) was also used as sensor for comparison. Thereafter, to investigate the omni-directivity of the $SH_0$ waves excited by the OSH-PT, seven face-shear $d_{24}$ PZT wafers with dimensions of $6 \text{ mm} \times 6 \text{ mm} \times 1.5 \text{ mm}$ were arranged around the OSH-PT transmitter at the intervals of 15° from 0° to 90° to serve as receivers. The layout and the locations of the transmitter and receivers are shown in Fig. 4(a). Finally, these seven face-shear $d_{24}$ PZT wafers were served as transmitters and the OSH-PT was used as receiver to explore the OSH-PT's performance on reception of $SH_0$ waves from all directions. All the transmitters were driven by a five-cycle Hanning window-modulated sinusoid toneburst signal provided by a function generator (33220A, Agilent, USA). A power amplifier (KH7602M) was used to amplify the drive signal and an Agilent DSO-X 3024A oscilloscope was used to collect the wave signals measured by receivers. In the received signals, the $SH_0$, $S_0$ and $A_0$ wave modes are identified based on their different group velocities. The group velocity dispersion curves of these three wave modes in the 1mm-thick aluminum plate are calculated and plotted in Fig. 4(b).

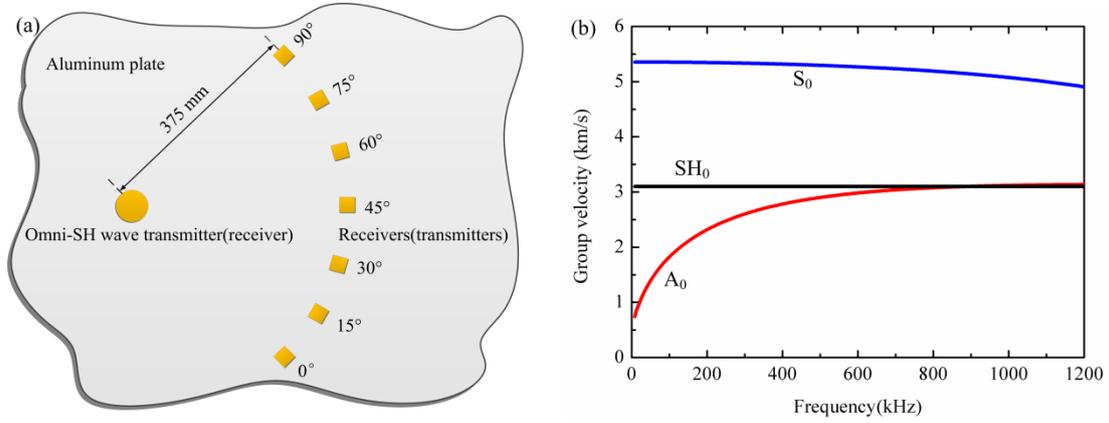

**Figure 4** (a) Schematic of the experimental setup to check the omni-directivity of the proposed OSH-PT. (b) Group velocity curves of the $SH_0$ wave and Lamb waves in a 1mm-thick aluminum plate.

4.1 *Excitation and reception of $SH_0$ wave by OSH-PT*

Fig. 5(a) shows the wave signals generated by the proposed OSH-PT and measured by a $d_{36}$ face-shear PMN-PT receiver placed on an arbitrary direction. The amplitude of the drive signal was fixed at 15 V and its central frequency was firstly set to be 190 kHz. As expected, $SH_0$ wave with high SNR (signal-to-noise) was successfully excited by the OSH-PT, as shown in Fig. 5(a). Furthermore, no other unwanted wave modes (such as $S_0$ and $A_0$ wave modes) were detected by the $d_{36}$ PMN-PT receiver. After analyzing these signals by using the continuous wavelet transform (CWT), the time interval of 123.95 μs can be extracted between the transmitter and receiver, as shown in Fig. 5(b). Bearing in mind the distance between the transmitter and receiver is 375 mm, the obtained group velocity of 3025 m/s is in good agreement with the theoretical group velocity (3099 m/s) of $SH_0$ wave in this aluminum plate.

As mentioned in Section 2, the maximum wave signals are expected to be obtained when the half wavelength of the excited $SH_0$ wave equals to the height "ae" of the trapezoidal PZT elements used in the OSH-PT. In order to confirm the validity of the proposed OSH-PT size selection scheme, we vary the driving frequency from 100 kHz to 300 kHz with the driving voltage fixed at 15 V. The received peak-to-peak voltage were extracted from the $SH_0$ wave signals measured by the $d_{36}$ PMN-PT receiver and normalized against the maximum values, as show in Fig. 5(c). When the exciting central frequency varies from 100 kHz to 195 kHz, the amplitude of the obtained $SH_0$ wave increases monotonically and reaches its maxima at 195 kHz. It should be noted that the amplitude of the $SH_0$ wave can keep over 97% of the maxima when the excited frequency varies from 190 kHz to 215 kHz. In addition, when the driving frequency varies from 215 kHz to 300 kHz, the amplitude of the generated $SH_0$ wave decreases monotonically. In the frequency range from 190 kHz to 215 kHz, the optimized half wavelength of the SH0 wave varies from about 7 mm to 8 mm, which is very close to the designed height "ae" (8 mm) of the trapezoidal PZT elements used in the OSH-PT. This phenomenon again confirmed that the size design of the OSH-PT is effective. Similar phenomenon was also observed by previous researchers[13, 25].

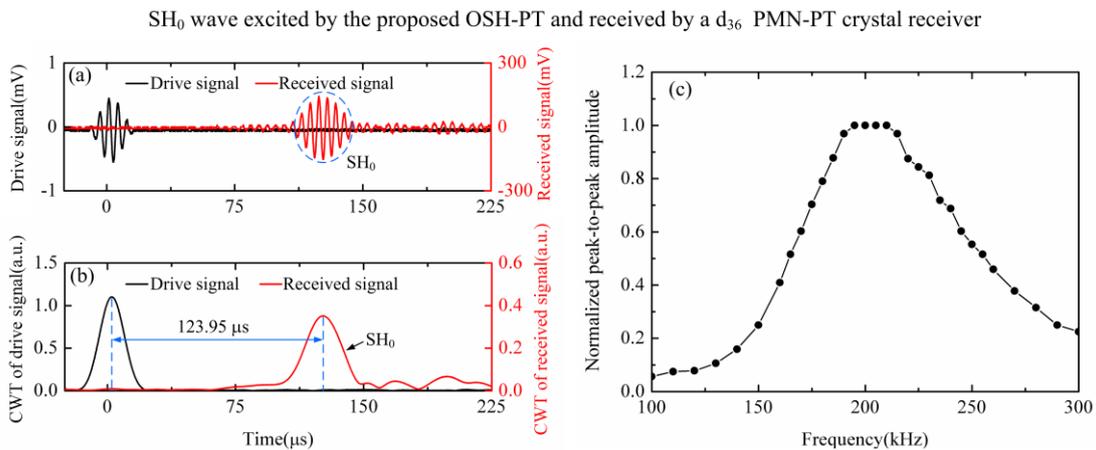

**Figure 5** (a) Driving signals applied to the OSH-PT at 190 kHz and received signals by a $d_{36}$ PMN-PT crystal receiver. (b) Continuous wavelet transform (CWT) of the drive and received wave signals shown in Fig. 5(a). (c) Normalized peak-to-peak amplitude versus frequency curve of the $SH_0$ wave generated by the OSH-PT.

Fig. 6(a) shows the $SH_0$ wave signals excited by the face-shear $d_{24}$ PZT wafer under a voltage of 24 V at 190 kHz and received by the proposed OSH-PT. For comparison, a $d_{36}$ PMN-PT crystal receiver was also used to detect the wave signals and the results were shown in Fig. 6(b). It can be seen from Fig. 6(a) that the excited $SH_0$ wave was successfully detected by the OSH-PT. Moreover, the waveform received by the OSH-PT is in good agreement with that received by the $d_{36}$ PMN-PT crystal receiver. This indicates that the proposed OSH-PT can be also used as a SH wave receiver.

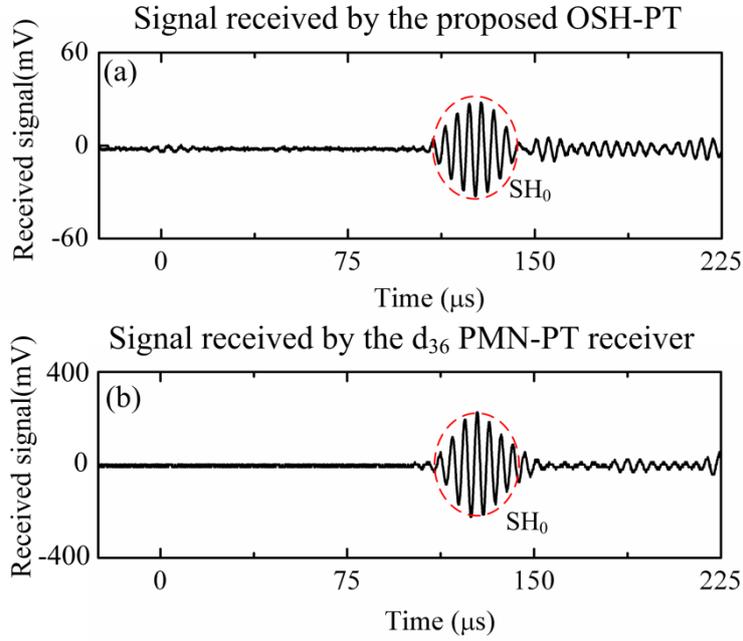

**Figure 6** SH0 wave signals excited by a face-shear $d_{24}$ PZT wafer under a voltage of 24V at 190 kHz and received by (a) the proposed OSH-PT and (b) a $d_{36}$ PMN-PT receiver.

4.2 *Experimental verification of omni-directivity of the OSH-PT*

Fig. 7 presents the amplitude directivity of the $SH_0$ waves excited by the proposed OSH-PT under a voltage of 15 V and received by seven distributed $d_{24}$ PZT receivers. As Fig. 5(c) indicates that best performance of the OSH-PT can be obtained in the frequency range from 190 kHz to 220 kHz, here the central frequency of the driving signals was selected at 190 kHz, 205 kHz and 220 kHz, respectively. The peak-to-peak amplitudes of the excited $SH_0$ waves were extracted and normalized with respect to the maximum amplitude of excited $SH_0$ wave at 205 kHz. As shown in Fig.7, at all the three frequencies, although the amplitude directivities of the excited $SH_0$ waves are not perfectly uniform across the angular range of 90°, the maximum deviation from the mean amplitude is relative small, i.e., 12.4% at 190kHz, 7.7% at 205kHz and 10.3% at 220kHz. The directional deviation may be attributed to the non-uniform bond layer between the trapezoidal PZT elements and the aluminum plate. The non-uniform bond layer will affect the shear resonance frequency of each individual PZT element and it will also have influences on the transfer of the shear deformation, leading to the disturbance of the excited $SH_0$ waves along different directions. Note that the non-uniformity of the bond layer can be avoided by improved fabrication processes. On the other hand, the directional deviation may be also caused by the non-uniform bond layer between the face-shear $d_{24}$ PZT wafer receivers and the aluminum plate. It is rather difficult to ensure that all the bonded receivers have the same sensitivity to $SH_0$ waves.

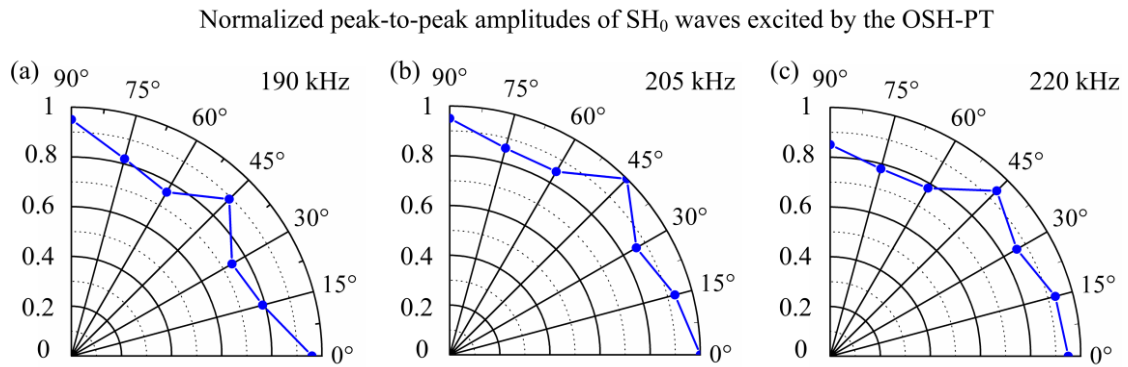

**Figure 7** The amplitude directivity of the $SH_0$ waves excited by the proposed OSH-PT at 190 kHz, 205 kHz and 220 kHz and received by angular distributed $d_{24}$ PZT receivers. The plots are normalized against the maximum amplitude of the excited $SH_0$ wave at 205 kHz.

Then the face-shear $d_{24}$ wafers placed at every 15° from 0° to 90° were used as transmitters to check the OSH-PT's performance on receiving the $SH_0$ waves in all directions. Similarly, the drive frequency was selected at 190 kHz, 205 kHz and 220 kHz, while the drive amplitude was fixed at 25 V. By extracting the peak-to-peak amplitudes of the $SH_0$ waves received by the OSH-PT, the amplitude directivity can be obtained, as shown in Fig. 8 where the plots were also normalized against the maximum amplitude of the received $SH_0$ wave at 205 kHz. It can be seen that for all the seven directions, the largest amplitude deviation from the mean peak-to-peak value is 12.1% at 190kHz, 9.8 % at 205 kHz and 12.3 % at 220 kHz, in a similar manner with the case that the OSH-PT serves as the transmitter. Based on Fig. 7 and Fig. 8, we can see that the proposed OSH-PT exhibits good omni-directivity properties, on matter it was used as a SH wave transmitter or a receiver.

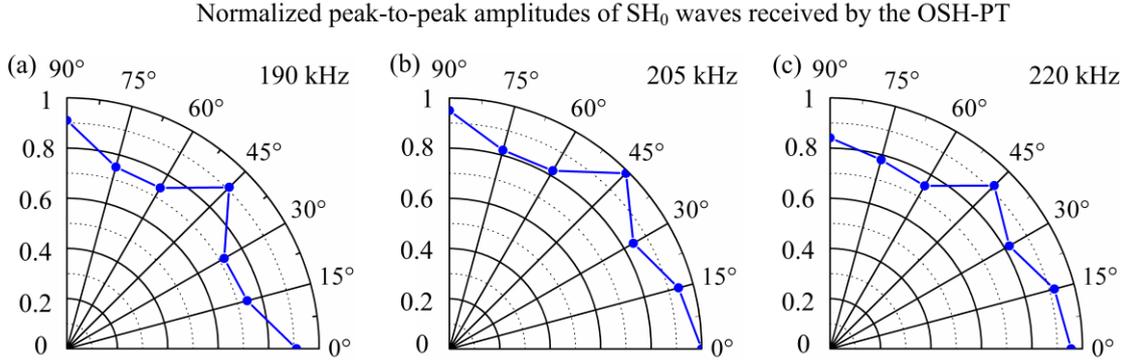

**Figure 8** The amplitude directivity of the $SH_0$ waves excited by angular distributed $d_{24}$ PZT transmitters and received by the proposed OSH-PT at 190 kHz, 205 kHz and 220 kHz. The plots are normalized against the maximum amplitude of the received $SH_0$ wave at 205 kHz.

## 5. Conclusions

In summary, an omnidirectional SH wave piezoelectric transducer (OSH-PT) was proposed which consists of a ring array of twelve face-shear $d_{24}$ trapezoidal PZT elements. Each PZT element can produce face-shear deformation under voltage, thus an effective circumferential shear deformation can be obtained in the proposed OSH-PT. The performances of the OSH-PT were examined by both finite element simulations and experimental testing. It was found that the OSH-PT can excite and receive the pure SH0 waves in a selected frequency range where no other unwanted wave modes such as Lamb wave were generated. The omni-directivity of the OSH-PT is proved to be fairly good whether it acts as a SH wave transmitter or receiver.

The OSH-PT proposed in this work may greatly promote the applications of SH wave based techniques in NDT and SHM. It can be used as the basic transducer for defect inspection in large plate-like structures, or as a transducer unit in the phased array system. The size of the OSH-PT can be designed and tailored according to the practically used frequency range. The low cost and compact size of the proposed OSH-PT makes it also suitable as sensors used in SHM. Moreover, as the SH wave in plates is analogue to the T(0,1) torsional wave in pipes, this work may also shed some light on the advanced NDT/SHM schemes in pipe-like structures.